# "Analysis of trap spectra in LEC and epitaxial GaAs"

(Invited talk)

J. Vaitkus, E.Gaubas, V.Kazukauskas, V.Rinkevicius, J.Storasta, R.Tomasiunas

*Institute of Materials Science and Applied Research & Semiconductor Physics Department,
Vilnius University, Vilnius, Lithuania*

K.M.Smith, V.O'Shea

*Department of Physics and Astronomy, Glasgow University, Glasgow, UK*

**Abstract**

Different methods of trap parameter measurement are analysed. Transient photoconductivity and thermally stimulated effects were used to investigate the influence of traps in LEC SI-GaAs and high resistivity epitaxial GaAs. The peculiarities of the TSC were analysed and shown to be related to the influence of crystal micro-inhomogeneities.

## Introduction

Investigations of deep centres in semiconductors pursue a number of different goals: 1) the understanding of non-equilibrium carrier lifetime and response to carrier concentration transients are important for definition of device parameters and operating regimes; 2) the characteristic parameters of deep levels contain information about the nature of defects and can be used in regulating material and device technology; 3) the behaviour of semiconductors during investigation of deep levels contains information about the material microstructure. The last of these was the main focus of this work. In general it creates a possibility to correct a model of material or device. The traps in SI-GaAs have been investigated in many papers [1-3 etc.] but not enough attention has been paid to the influence of inhomogeneity. In this work the analysis of non-equilibrium phenomena in semi-insulating and high resistivity GaAs crystals is shown to provide a method for qualitative evaluation of material inhomogeneity.

The most popular methods used to investigate deep centre parameters are related to studies of non-equilibrium conductivity or centre filling. In this work we review some peculiarities of methods based on optically excited photoconductivity / non-equilibrium free carrier concentration transient behaviour after light pulse excitation and / or thermal stimulation of conductivity and their application to investigation of GaAs. These phenomena can be used to measure the trap parameters, (capture cross-section, activation energy and concentration of empty traps can be extracted from TSC data), which can then be used to calculate the lifetime for capture into the trap(s) and the re-trapping time. A comparison of these calculated parameters and the non-equilibrium carrier concentration decay time allows a deeper understanding of the non-equilibrium carrier behaviour after any type of excitation.

The investigation of compensated GaAs has some difficulties if the crystal is semi-insulating or of very high resistivity material. The main problems, (which in many cases are not taken into account), are related to the micro-inhomogeneities in wafers. The importance of particular TSC dependencies for crystal quality evaluation is pointed out

## Experimental results and discussion

**A.** Investigation of transient photoconductivity / non-equilibrium free carrier concentration transient behaviour after light pulse excitation.

This investigation was performed by measurement of microwave absorption or refractive index transient behaviour following excitation by a short light pulse. These measurements depend on the non-equilibrium carrier concentration time dependence. The initial and final parts of this dependence allow the measurement of the lifetime and the evaluation of the influence of re-trapping [4]. The initial part of the carrier concentration decay gives information about carrier lifetime. It was measured by means of the light-induced free carrier grating method [5], through which the non-equilibrium free carrier concentration decay is measured from the time variation of the diffraction efficiency $\eta$, ($\Delta n \sim \sqrt{\eta}$), of a grating excited by the interference of light beams from a short light pulse and sampled by a delayed fraction of the same beam. It was established that the lifetime of carriers in LEC GaAs was approximately (2.4±0.7) ns and in epitaxial GaAs – (6±1) ns. The intensity for the excitation was chosen in the range of linear generation of carriers that appears in a two step generation of free carriers. At higher intensities some peculiarities of saturation appear that will be analysed elsewhere (Fig.1). The final stages of non-equilibrium carrier decay demonstrate the influence of carrier re-trapping. This part of the dependence, measured by a microwave absorption method [5], reveals that re-trapping effects in typical high quality LEC (Hitachi) and high resistivity epitaxial (Aixtron) GaAs cause the momentum time constant in the main part of the decay to be of the order of a few hundreds of nanoseconds. A slower component, with time constant equal to some microseconds, was also found to exert a small influence (Fig.2).

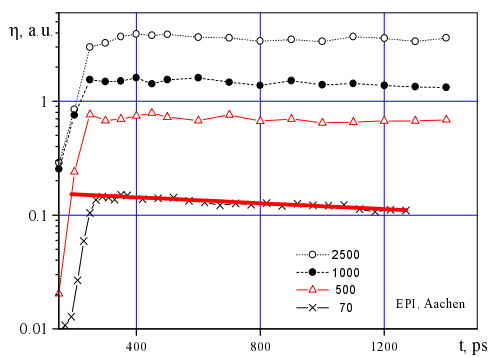
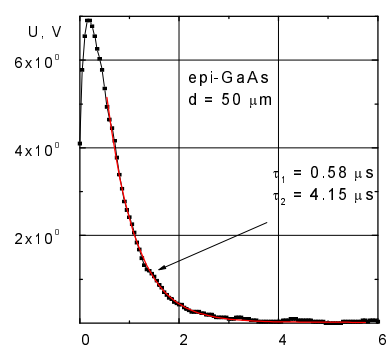

Fig.1. Decay of diffraction grating efficiency, $\eta$, in epi-GaAs after excitation by a short light pulse. The crosses, open triangles, closed and open circles correspond to increasing intensities of excitation, (the crosses to the limiting intensity for linear generation).

Fig.2. Final part of free carrier concentration ($\Delta n \propto U$) decay after short light pulse excitation.

**B.** Most simple evidence of deep levels is obtained from thermally stimulated effects. One of the "oldest" methods of this type is the investigation of thermally stimulated conductivity. This effect is very simple for homogeneous material with rather simple trap spectra. Then level parameters such as activation energy, carrier capture cross-section and concentration can be determined. The case of high resistivity GaAs is more complicated than a "classical" semiconductor with traps and it requires a very careful and detailed analysis of TSC spectra.

In contributions to previous workshops [6,7] we have demonstrated the peculiarities of TSC in GaAs detectors. It was shown that the main deviation from classical TSC is related to the existence of crystal inhomogeneities. For the present measurements, the epitaxial layer was grown on thinned standard, semi-insulating substrate material and ohmic contacts were deposited on the top of the epi-layer. The analysis of crystals fabricated in traditional bar shapes, (approx. 1.5 mm long, 1mm wide, 100µm thick), showed behaviour very similar to the Schottky detectors but this sample geometry allowed a more straightforward analysis of results. Some aspects of the TSC measurements, however, appear to exhibit a residual influence of polarisation of the substrate in the measurements. It was assumed that re-trapping is very small. In this case it is rather easy to simulate TSC spectra if the activation energies are determined [8]. Activation energies were measured by TSC using the multiple heating cycle method. The initial part of the TSC was corrected to achieve an exponential dependence by compensation of polarisation induced current. The coincidence of experimental and simulated TSC allows the estimation of the trapping centre parameters. (Examples are given in Fig.3 and Table 1). The simulation of multiple heating TSC with these parameters does not give such good agreement between experimental and simulation results. This disagreement appears due to the difficulty of excluding the influence of crystal inhomogeneity. Paradoxically, better correspondence of experimental and simulation results was achieved in highly irradiated samples. This could be explained on the basis that irradiation defects create on average more homogeneous media.

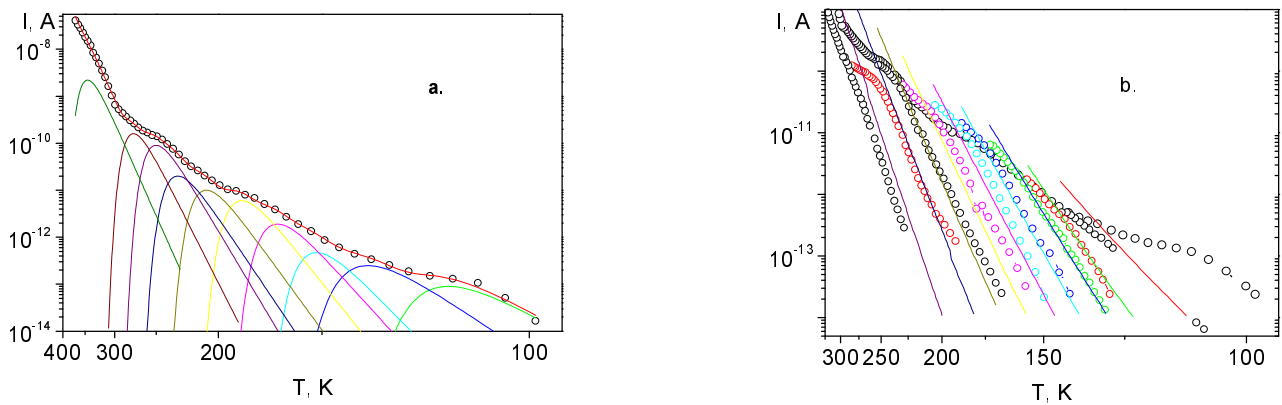

Fig.3. TSC after short pulse excitation in high resistivity epitaxial GaAs. a – heating after short pulse excitation; b – TSC in multiple heating regime. Points – experiment, lines – simulation.

**Table 1**
Trap parameters for GaAs Epi-1 sample (L=0.15 cm, S=0.001 cm$^2$, U=4V)

| $\Delta E_M$, meV | $\gamma_{nM}$, cm$^3$/s | $\tau_{retrapping}$, s | (M-m$_0$), cm$^{-3}$ | $\tau_{capture\ min}$, s | $T_{max}$, K | $I_{max}$, A |
|---|---|---|---|---|---|---|
| 145 | 8.7E-13 | 0.16E-3 | 5.72E+11 | 2.0 | 115 | 9.00E-14 |
| 175 | 8.3E-12 | 0.25E-3 | 1.93E+12 | 0.63 | 135 | 2.50E-13 |
| 270 | 2.1E-10 | 0.39E-3 | 3.07E+12 | 1.6E-3 | 152 | 4.70E-13 |
| 290 | 8.4E-11 | 2.1E-3 | 1.41E+13 | 0.84E-3 | 168 | 1.90E-12 |
| 330 | 1.1E-10 | 7.4E-3 | 4.80E+13 | 0.19E-3 | 186 | 6.00E-12 |
| 360 | 4.7E-11 | 0.055 | 9.16E+13 | 0.23E-3 | 208 | 1.00E-11 |
| 390 | 2.3E-11 | 2.8 | 2.07E+14 | 0.21E-3 | 230 | 2.00E-11 |
| 440 | 3.9E-11 | 1.4 | 9.76E+14 | 26.5E-6 | 250 | 9.00E-11 |
| 520 | 1.4E-10 | 8.7 | 1.79E+15 | 3.9E-6 | 275 | 1.60E-10 |
| 590 | 8.0E-12 | 2200 | 3.38E+14 | 0.37E-3 | 345 | 2.20E-09 |
| | | | | 1/($\Sigma$ 1/$\tau_{capture\ min}$) = | 3.2E-6 | |

$\tau_{capture\ min}$ = 1/ $\gamma_{nM}$ (M-m$_0$), $\tau_{retrapping}$ = 1/ $\gamma_{nM}$ N$_c$ exp(-$\Delta E_M$ /kT), $\gamma_{nM}$ = v$_{th}$ $\sigma_n$ is the recombination coefficient, v$_{th}$ – thermal velocity of electrons, $\sigma_n$ - electron capture cross-section; at T=300K v$_{th}$= 4.3E7 cm/s, N$_c$=4.0E17, cm$^{-3}$

Additional indirect evidence of the existence of a drift barrier follows from the comparison of quantitative data on the measured lifetime and the time constant calculated from TSC data. The calculated values are too small in comparison with experimental data. This can be explained as a drift barrier influence that reduces the absolute value of TSC. It gives a trap concentration which is smaller than the true value and increases the magnitude of the capture time constant. A short carrier lifetime causes a smaller value of TSC in comparison with the value if all carriers were extracted into the contacts. This results in a capture cross-section which is smaller than the true value.

The rather large number of TSC characteristic energies suggests that these energies do not always correspond to different deep centres. Due to the existence of micro-inhomogeneities the measured activation energy includes the trap ionisation energy and barrier height if this barrier surrounds the trap. Thermally generated free carriers can change these drift barriers. Then after the next heating cycle the activation energy will be different from the previous measurement.

Modulation of the local electric field, polarisation of the sample and barrier creation are illustrated by TSC measurements with different excitation conditions (Fig.4). If the (unbiased) sample was excited by 1.06 eV light for 1s, and after excitation it was kept for 5 minutes in the dark, then the TSC shows only a few clearly pronounced peaks (Fig.4, curve 1). If the same measurement was performed with the sample biased then the low temperature part became smoother and the different trap activation regions were difficult to separate (Fig.4, curve 2).

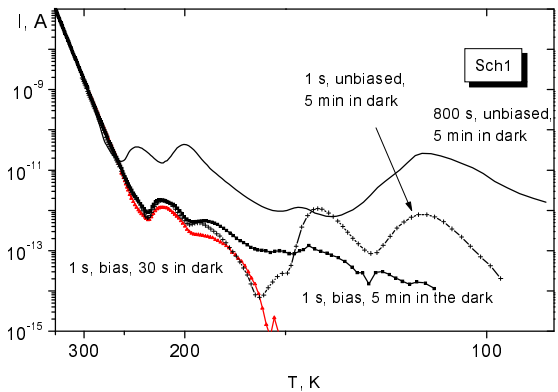
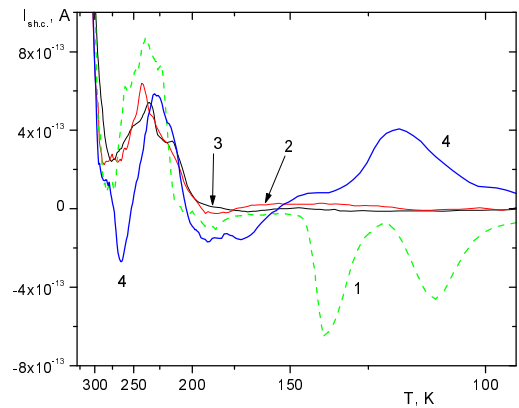

Fig.4. Temperature dependence of (a) TSC and (b) thermal depolarisation short circuit current in LEC SI-GaAs Sch1 sample after different excitation regimes: (1) – excitation for 1 s, no bias, 5 min in dark; (2) – 1 s, bias on , 30 s; (3) – 1s, bias on , 5 min; (4) – 800 s, no bias, 5 min, respectively.

This may be explained in terms of the influence of drift barriers in separating thermally ionised carriers due to percolation from recombination centres by a potential barrier that suppresses the recombination of carriers. Furthermore, repeating the same measurement but keeping the sample in the dark for a shorter time, (only 30 s), showed that, at low temperature, the sample was polarised. The TSC flowed in the opposite direction and the "normal" TSC appeared only in the region T ≥ 160 K. A coincidence of the TSC current in this region with the TSC in an unbiased sample suggests that the minimum at T=160 K on curve 1 is also related with polarisation of the sample. If the excitation was long then the EL2 centres were bleached, the sample changed conductivity to p-type and this caused a significant change of TSC behaviour at low temperature.

Such a complicated behaviour of drift barriers was also seen in measurements of thermally stimulated depolarisation (TSD) (Fig.4(b)). Varying the excitation duration produces different modulation of the electric field in the sample and thermally activated carriers recharge the sample in the short circuit regime depending on the polarisation polarity at a certain temperature. Different signs of TSD maxima show that there are various barrier types in the bulk of a sample and there is an overlap of the depolarisation curve with the thermopower effect at high temperature. This could influence the depolarisation current at lower temperature, too, but to separate these effects is rather difficult.

From the analysis of different types of SI-GaAs it is possible to establish the influence of inhomogeneities related to drift barriers and to try to separate the "real" and "effective" capture centres from TSC spectra. As an important informative detail of TSC we see a monotonic change of dependence. The change can be related to two processes: a) a standard phenomenon in TSC – the emptying of one centre and the appearance of thermal stimulation of a deeper centre; or b) thermal activation of minority carriers that changes the lifetime and the drift barrier. In the latter case it is possible, (as was used for photo-compensation of SI-GaAs in [9]), for a change of TSC to occur due to restoration of $EL2^0$ from $EL2^{**}$ and subsequent transformation of $EL2^0$ to $EL2^+$ by capture of a thermally excited free hole. The measurement of the TSC activation energy in the multiple heating regime is a key to the explanation of the nature of the TSC. A change of activation energy in the region where one type of centre is emptying is connected with a drift barrier modulation and lifetime change due to recombination barrier temperature activation or temperature dependence of the capture cross-section.

The proposed analysis of the initial part and activation energy of the TSC transforms this method of trap parameter measurement into a method for the identification of electrically active inhomogeneities. The characterisation of material according to the inhomogeneity type is important for any type radiation detector and it is a very sensitive tool for establishing the thresholds of degradation events in crystal and detector.

Comparison of the detector and crystal results provides the possibility of more detail than obtainable by other methods such as analysis of non-equilibrium current percolation and charge collection after pulse excitation. The clearly expressed influence of inhomogeneities on carrier percolation current gives a supplementary model to explain the decrease of lifetime in the regions of high electric field [10]. If EL2 centres are concentrated around the dislocations and a local space charge exists in these regions, then the recombination through EL2 centres will be more efficient if electrons have excess energy. It should be noted that this local electric charge creates a field that attracts holes to be captured by neutral EL2. This process also increases the recombination efficiency. The nature of the internal electric field modulation is not yet clear, but crystal tension and compression around the dislocation is a good starting point. Related electro-optical effects have been observed [11] and the influence of traps on the refractive index in the optical region has also been investigated [12]. The photo-voltage maximum at low temperature [9] was observed at a slightly higher temperature than the typical temperature at which the modified EL2 centre returns to a normal state. In the same temperature region, redistribution of the electric field in high quality LEC crystals has been observed [11]. These data show that these local field regions create inhomogeneities of crystal electrical properties that can be reduced after heating to above 420K. As this electrical inhomogeneity of the material is dependent on non-equilibrium carriers it is important to reduce it in the material for radiation detectors, especially for those that operate below room temperature.

## Conclusion

The investigation of trap spectra by thermally stimulated phenomena reveals micro-inhomogeneities in GaAs crystals and detectors. The influence of these inhomogeneities on the behaviour of semi-insulating LEC and high resistivity epitaxial GaAs detector is supported by a comparison of measured non-equilibrium carrier concentration decay time constants with those calculated from trap parameters determined from TSC.

## Acknowledgements

This work was supported by The Royal Society long-term project and a Vilnius University Science Fund grant.